\documentclass[review]{elsarticle}

\usepackage{lineno,hyperref}

\journal{Journal of Parallel and Distributed Computing}

\begin{document}

\begin{frontmatter}

\title{Soft Error Resilience and Failure Recovery for Continuum Dynamics Applications\tnoteref{mytitlenote}}
\tnotetext[mytitlenote]{This work was performed at the Ultrascale Systems Research Center (USRC) and the Applied Computer Science Group (CCS-7) at Los Alamos National Laboratory (LANL), assigned the LANL identifier LA-UR-17-?????.}


\author[mymainaddress]{Li Tan\corref{mycorrespondingauthor}}
\cortext[mycorrespondingauthor]{Corresponding author}
\ead{darkwhite29@gmail.com}

\author[mysecondaryaddress]{Marc Charest}

\author[mymainaddress]{Nathan DeBardeleben}

\author[mymainaddress]{Qiang Guan}

\author[mysecondaryaddress]{Ben Bergen}

\address[mymainaddress]{Ultrascale Systems Research Center, Los Alamos National Laboratory, Los Alamos, New Mexico, USA}
\address[mysecondaryaddress]{Applied Computer Science Group, Los Alamos National Laboratory, Los Alamos, New Mexico, USA}

\begin{abstract}
The persistently growing resilience concerns of large-scale computing systems today require not only generic fault tolerance approaches, but also application-level resilience, due to demanding efficiency and various domain-specific requirements. Scientific applications within a particular domain generally comply with domain conservation laws, which can be leveraged as an error detection criterion to study the resilience of this domain of applications sharing similar program characteristics. However, it is challenging to achieve application resilience: (a) how to identify the invariants of a given domain of applications, knowing the conservation laws, and (b) how to utilize the invariants to efficiently detect and recover from failures in application runs.

In this work, we target several continuum dynamics software packages, FleCSALE \cite{flecsale} and CODY \cite{cody} (with intrinsic invariants during computation), study their resilience to soft errors online (injected using an open-source fault injector), and investigate the opportunities for non-intrusive and lightweight failure recovery (checksum-based invariant checking). We propose a checksum-retry approach to achieve our goals, and experimental results on a virtualized platform with extensive fault injection campaigns demonstrate the effectiveness and efficiency of the proposed approach.
\end{abstract}


\end{frontmatter}


\section{Introduction}

As High Performance Computing (HPC) technology advances, large-scale computing systems today with numerous compute nodes suffer from a growing number of hard and soft errors at an increasing rate that lead to highly expensive system costs. Although generic resilience techniques like Checkpoint/Restart (C/R) and Triple Modular Redundancy (TMR) have been widely applied in HPC environments with high generality, scalability, and efficiency \cite{sc10}, application resilience remains to be of great concern due to specific domain requirements such as computational accuracy and time sensitivity \cite{pact12}. Due to the nature of targeting applications sharing common program characteristics (e.g., computing patterns, similar data distribution, and similar deadline/requirements of program outputs), application-specific resilience methods generally benefit from performance efficiency and high resilience, compared to generic resilience techniques. On the contrary, they have the challenges of being general-purpose to more types of applications.

Application resilience requires well-utilized application-specific knowledge to better design fault tolerance approaches, such as Algorithm Based Fault Tolerance (ABFT) for numerical linear algebra operations. Widely employed in different fields, domain-specific scientific applications in general comply with domain conservation laws, e.g., checksum-based data redundancy in matrix operations for numerical linear algebra applications \cite{scalapack}, and conservation of momentum and conservation of energy for hydrodynamics \cite{clamr} and continuum dynamics applications \cite{flecsale} \cite{cody}. For applications conducting simulation of computational scenarios that follow these fundamental laws, we are able to leverage some invariants (which should always be constant during the computation according to the laws) to detect abnormality in program runs. Similar to Diskless Checkpointing (DC), for continuum dynamics applications, a retry model based on checksums of such invariants can effectively recover from soft errors online in a lightweight fashion.

FleCSALE \cite{flecsale} and CODY \cite{cody} are two suites of representative aforementioned domain-specific applications of concern by US Department of Energy (DOE), for studying multi-phase continuum dynamics problems with different runtime environments. It is critical to study and improve the resilience of such applications to meet the exascale vision in the next decade. In this work, we investigate the susceptibility of continuum dynamics programs to soft errors by utilizing the open-source fault injector F-SEFI \cite{ipdps14a} \cite{fsefi}, and evaluate our failure recovery approach based on checksum-retry. We demonstrate that due to the compute-intensive program characteristics, continuum dynamics applications are prone to soft errors, primarily Silent Data Corruption (SDC). We can effectively recover program runs from the injected soft errors by our checksum-retry approach with negligible performance loss. In summary, the contributions of this paper include:

\begin{itemize}
\item We study the resilience to soft errors of continuum dynamics applications based on program characteristics, and propose a checksum-retry approach for high performance failure recovery, which is implemented portably at library level for generality purposes;
\vspace{2mm}\item Soft error vulnerability and failure recovery of FleCSALE and CODY programs are experimentally evaluated on a virtualized platform using the F-SEFI fault injector. Emulated faulty runs can be effectively recovered with negligible 1.4\% performance loss on average.
\end{itemize}

The remainder of the paper is organized as follows: Background knowledge is briefly introduced in Section 2. Details of the resilience study of continuum dynamics applications are presented in Section 3. Section 4 provides experimental results. We discuss related work in Section 5, and Section 6 concludes.

\section{Background}

\subsection{Continuum Dynamics Applications}

Continuum Dynamics is a fundamental branch of mechanics that deals with the dynamical behavior of materials modeled as a continuous mass rather than as discrete particles, which is of great concern by US Department of Energy since the theory is representative of many important scientific workloads. FleCSALE is a software package consisting of several mini-applications for studying multi-phase continuum dynamics problems with different runtime environments. FleCSALE is built on top of the Flexible Computational Science Infrastructure (FleCSI) \cite{flecsi}, a compile-time configurable framework providing a set of reusable and efficient computational science infrastructure tools for implementation aid of multi-physics applications. Current support includes multi-dimensional mesh topology, mesh geometry, mesh adjacency information, n-dimensional hashed-tree data structures, graph partitioning interfaces, and dependency closures. CODY \cite{cody} is a suite of mini-application implementation, characteristic of continuum dynamics (e.g., Godunov hydrocode) using a number of standard HPC parallelization systems. The implementation is designed primarily to test the performance in different computer architectures, for example, when running a standard simple compressible fluid dynamics model. In this work, we select the core programs within FleCSALE and CODY as our target applications for a resilience study of the field of Continuum Dynamics.

\subsection{Register-level Fault Injector: F-SEFI}

Targeting HPC applications and leveraging the virtual machine infrastructure QEMU \cite{atc05}, F-SEFI was proposed as a fine-grained soft error fault injector for profiling software robustness against soft errors. By injecting soft errors at register level online, the impacts of real hardware faults occurring at logic circuits can be effectively represented and studied at software level for building resilient HPC applications. In a highly configurable user-customized manner, the fault injector is designed to control what application, which function, and when to inject soft errors into what instructions executed, in a non-intrusive fashion, i.e., during the fault injection using F-SEFI, there is neither required modification to source code, compiler, and the Operating System (OS), nor the need of extra hardware. It runs in the user space alongside HPC applications, intercepts instructions translated from the guest OS (where virtual machines runs) to the host OS (where hypervisor runs), and replaces them with the contaminated version before they are delivered to the host OS for execution. It has been publicly released \cite{fsefi} and widely employed in scientific application resilience studies \cite{cluster15} \cite{parco17}.

\section{Resilience Methodology: A Case Study of FleCSALE}

In this section, as a case study, we introduce the FleCSALE software package, a set of representative continuum dynamics applications. Firstly we analyze the source code of the applications by discussing the details of tasks fulfilled by the code and pinpointing vulnerable functions to soft errors. Secondly we propose our invariant-based failure recovery approach to detect and correct potential soft errors online.

\subsection{Source Code Organization and Structure}

FleCSALE is a scientific software package developed for studying problems that can be characterized using continuum dynamics \cite{flecsale}, e.g., fluid flow. It consists of two mini-applications as an initial stage of development: \emph{hydro} (an \emph{Eulerian} solver, i.e., the mesh is fixed and the fluid flows through it) and \emph{maire\_hydro} (a \emph{Lagrangian} solver, i.e. the mesh moves with the fluid). In particular, FleCSALE utilizes the Flexible Computational Science Infrastructure (FleCSI) \cite{flecsi} for mesh and data structure support. In general, both applications work as follows: Firstly a mesh is built and initialized together with its entities including points, cells, faces, edges, and vertices; secondly the connectivity between different mesh entities is computed using the \emph{Eulerian} solver (\emph{hydro}) and the \emph{Lagrangian} solver (\emph{maire\_hydro}) individually; lastly the total sum of system invariants (e.g., mass, momentum, and energy) are printed to standard output and the variables of individual entities during the computation (e.g., velocity, temperature, and pressure) are dumped into disk files.

At source code level, first of all some setup work is done for building and initializing the mesh, followed by a main loop that iterates throughout all mesh entities while doing the computation of entity connectivity. The users can specify the input size and the output frequency, as well as the output format. There exists several compute-intensive functions that finish solving the multi-phase system, e.g., \emph{evaluate\_fluxes}(), \emph{evaluate\_corner\_coef}(), \emph{evaluate\_nodal\_state}(), and \emph{evaluate\_forces}(). They are highly parallelizable (we parallelize them using OpenMP in the current release), and due to their intensive accesses to registers and memory, they consume the majority of the execution time and are prone to potential soft errors on mesh entities during the computation at scale.

\subsection{Invariant-based Error Detection and Recovery}

As a representative continuum dynamics application suite, FleCSALE produces computational connectivity between mesh entities, which follows its domain-specific conservation laws in general, specifically conservation of the total momentum and conservation of the total energy. Leveraging checksums of the invariants during computation, runtime errors can be effectively detected. Additionally, upon the detection of errors, we can correct some errors such as single-bit flips by rolling back the main loop to the last saved correct iteration, as the last saved correct application state, similar to the concept of Checkpoint/Restart (C/R). Therefore, we propose a checksum-based retry model for high performance error detection and correction in this scenario.

\begin{figure}[h]
\centering
\includegraphics[width=1.79in]{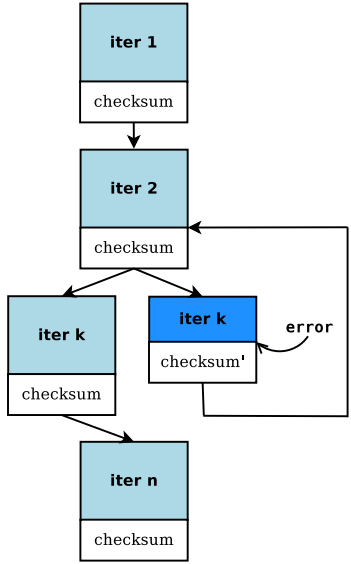}
\caption{The Checksum-Retry Model for Invariant-based Error Detection and Recovery.}
\label{checksum_retry}
\end{figure}

As shown in Figure \ref{checksum_retry}, when the main computational loop iterates, the calculation of checksums of all field variables of the mesh involved in the computation is added at the end of each iteration. We are interested in the checksums of the invariants, e.g., the sum of the total energy, since they should always be constant throughout the computation. If the bit-wise comparison between checksums of invariants at the $k$th iteration shows some discrepancy, it indicates that some errors occurred at the $k$th iteration that contaminated the values of invariants. Upon the detection of errors, we roll back the loop counter to the previous iteration and restore all values from previously saved correct ones (note that we only hold correct computation data in the last iteration in memory and overwrite it when a new correct iteration completes for space limitation), until the computation completes. This failure recovery model is lightweight because all recovery steps are performed in the memory online: No disk accesses and network communications are required, and thus no application restart is necessary. The model is also scalable in the sense of local checking: Each compute node has a copy of the global conserved invariants, and after each iteration, the newly-computed invariants will be compared with the shadow copy locally. If they vary beyond a certain tolerance value, the current computation results will be discarded by rolling back to the previous iteration, until the invariants match or the number of retries exceeds a threshold. That is, all the checking is done locally, and does not need global communication which may affect the scalability.

\section{Evaluation}

In this section, we present details of experimental evaluation on our resilience study on the core continuum dynamics applications within FleCSALE and CODY, a total of four mini-applications running on a virtualized platform within an HPC server. Overall, the empirical resilience study aims to showcase that: (a) continuum dynamics applications are susceptible to soft errors -- register-level fault injection in the compute-intensive functions of benchmarks has a considerable probability of corrupting the computation results, and (b) our checksum-based retry approach is capable of recovering from soft errors in this setup effectively and efficiently at runtime. We utilize the open-source fault injector, F-SEFI, to perform all fault injection campaigns. Specifically, for each benchmark, we conducted 2000 runs with one fault injected in each run for each interested instruction type, and collected the statistical failure data using self-coded post-processing scripts.

\subsection{Experimental Setup}

\begin{table}[h]
\scriptsize
\centering
\caption{Hardware Configuration for All Experiments.}
\label{hardware_configuration}
\begin{tabular}{|c|c|}
\hline
System Size & 40 logical cores\\
\hline
Processor & Intel Xeon E5-2660 (10-core, hyper-threading)\\
\hline
CPU Frequency & 1.2 to 2.6 GHz incremented by 0.1 GHz\\
\hline
Memory & 128 GB RAM\\
\hline
Cache & 640 KB L1, 2560 KB L2, 25600 KB L3\\
\hline
OS & Ubuntu 14.10, 64-bit Linux kernel 3.16.0\\
\hline
\end{tabular}
\normalsize
\end{table}

Table \ref{hardware_configuration} lists the hardware configuration of the experimental HPC server equipped with two Intel Xeon E5-2660 processors. We leverage F-SEFI to establish a virtualized Linux environment as a guest node running on top of the host node configured using the specification in Table \ref{hardware_configuration}. Detailed steps of fault injection proceed as follows: Booting from a disk image with pre-installed OS and software, the guest node launches the original runs of FleCSALE and CODY, and the hypervisor of the guest virtual machine intercepts and translates each instruction of FleCSALE and CODY runs on the guest node, into an instruction to be executed on the host architecture. During the translation, corrupted values are injected by flipping bits in the registers used in the program execution.

\subsection{Empirical Implementation}

For generality purposes, we implemented our invariant-based checksum-retry approach at library level, i.e., the module performing checksum-retry is organized as a library called by the evaluated applications. Specifically, the checksum-retry module consists of two parts: (a) a checksum utility of calculating the checksum values of the output solution data (e.g., all field variables). The checksum code computes an MD5 checksum (a 128-bit hash value) and put it into a predefined local variable, which has string and byte representations of the sum. Other digests could be specified although it is defaulted to be MD5 (and tested to be sufficient in our experiments), and (b) a retry utility that firstly compares the checksum values of invariants across loop iterations as discrepancy checking, and next rolls back to the last correct loop iteration with saved computation values restored, if checksum discrepancy is detected. Note that the rolling back functionality will not be triggered if checksums remain the same (the checksum checking overhead is highly negligible and thus only the total overhead of checking and recovery is reported in the later text). We limit the number of retries to a user-customized constant for not incurring an infinite loop of failure recovery if errors re-occur in restored iterations.

We provide highly encapsulated APIs to achieve transparent failure detection and recovery in a black-box fashion. The users of our approach do not need to know what types of soft errors and when the errors occur in their applications. The parameters need to pass to the APIs are basically the interested function names (the \emph{main}() function by default if the users have no access to the source code). Our approach considers the resilience of all instructions executed by users' applications. We also offer the users to specify the fault tolerance threshold values for their applications, i.e., a difference in output defined based on the physics or any internal theories. Users can configure the threshold values based on the real problems they want to solve, or different machines they are running the applications on that produce different machine epsilon. Typical threshold values can be either some numbers of the order of machine epsilon (the difference in value), or of massive output errors (the difference in magnitude).

\subsection{Soft Error Vulnerability}

\subsubsection{Failure Distribution}

\begin{table}[h]
\scriptsize
\centering
\caption{Failure Distribution of Fault Injection into Four Continuum Dynamics Applications.}
\label{failure_distribution}
\begin{tabular}{|c|c|c|c|c|}
\hline
 & \emph{hydro} & \emph{maire\_hydro} & \emph{mish} & \emph{umma}\\
\hline
\texttt{fadd} & SDC & SDC & SDC & SDC\\
\hline
\texttt{fmul} & SDC & SDC & SDC & SDC\\
\hline
\texttt{cmp} & SDC & crash & SDC & SDC\\
\hline
\texttt{imul} & N/A & Assertion Failure & crash & N/A\\
\hline
\texttt{xor} & benign & N/A & benign & N/A\\
\hline
\end{tabular}
\normalsize
\end{table}

Table \ref{failure_distribution} shows the distribution of injected faults into four mini-applications within FleCSALE (\emph{hydro} and \emph{maire\_hydro}) and CODY (\emph{mish} and \emph{umma}), when fault injection was conducted using F-SEFI. We can see that both \texttt{fadd} and \texttt{fmul} instructions are susceptible to SDC in different local/global parameters, i.e., contaminated values of program outputs. While injecting faults into \texttt{cmp} instructions in \emph{hydro}, \emph{mish}, and \emph{umma} runs causes SDC, corrupting \texttt{cmp} of \emph{maire\_hydro} causes crashes (segmentation faults) only. This is because the interested \texttt{cmp} instructions of \emph{hydro}, \emph{mish}, and \emph{umma} are used for computation purposes (comparing data only), but the \texttt{cmp} instructions of \emph{maire\_hydro} are employed in branching (addressing), i.e., corrupted pointers from fault injection account for crashes due to illegal memory accesses. Note that \texttt{imul} instructions are only used in \emph{maire\_hydro} and \emph{mish} but not in \emph{hydro} and \emph{umma}. \texttt{xor} instructions are only employed in \emph{hydro} and \emph{mish} but not in \emph{maire\_hydro} and \emph{umma}. However, there are no observable errors (i.e., benign) when injecting faults into \texttt{xor} instructions in runs of \emph{hydro} and \emph{mish}.

\subsubsection{Failure Rates}

\begin{table}[h]
\scriptsize
\centering
\caption{Failure Rates of FleCSALE Application \emph{hydro}.}
\label{failure_rate_hydro}
\begin{tabular}{|c|c|c|}
\hline
\emph{Instruction} & \emph{Failure Type} & \emph{Failure Rate}\\
\hline
\texttt{fadd} & SDC in \emph{fluxes} & 1189/2000 (59.5\%)\\
\hline
\texttt{fmul} & SDC in \emph{fluxes} & 1277/2000 (63.9\%)\\
\hline
\texttt{cmp} & SDC in \emph{fluxes} & 702/2000 (35.1\%)\\
\hline
\end{tabular}
\normalsize
\end{table}

\begin{table}[h]
\scriptsize
\centering
\caption{Failure Rates of FleCSALE Application \emph{maire\_hydro}.}
\label{failure_rate_mhydro}
\begin{tabular}{|c|c|c|}
\hline
\emph{Instruction} & \emph{Failure Type} & \emph{Failure Rate}\\
\hline
\texttt{fadd} & SDC in \emph{momentum} & 1151/2000 (57.6\%)\\
\hline
\texttt{fmul} & SDC in \emph{momentum} & 1544/2000 (77.2\%)\\
\hline
\texttt{cmp} & crash (Segmentation Fault) & 908/2000 (45.4\%)\\
\hline
\texttt{imul} & Assertion Failure in \emph{energy} & 729/2000 (36.5\%)\\
\hline
\end{tabular}
\normalsize
\end{table}

\begin{table}[h]
\scriptsize
\centering
\caption{Failure Rates of CODY Application \emph{mish}.}
\label{failure_rate_mish}
\begin{tabular}{|c|c|c|}
\hline
\emph{Instruction} & \emph{Failure Type} & \emph{Failure Rate}\\
\hline
\texttt{fadd} & SDC in \emph{fluxes} & 874/2000 (43.7\%)\\
\hline
\texttt{fmul} & SDC in \emph{fluxes} & 1027/2000 (51.4\%)\\
\hline
\texttt{cmp} & SDC in \emph{fluxes} & 1106/2000 (55.3\%)\\
\hline
\texttt{imul} & crash (Segmentation Fault) & 563/2000 (28.2\%)\\
\hline
\end{tabular}
\normalsize
\end{table}

\begin{table}[h]
\scriptsize
\centering
\caption{Failure Rates of CODY Application \emph{umma}.}
\label{failure_rate_umma}
\begin{tabular}{|c|c|c|}
\hline
\emph{Instruction} & \emph{Failure Type} & \emph{Failure Rate}\\
\hline
\texttt{fadd} & SDC in \emph{edges} & 691/2000 (34.6\%)\\
\hline
\texttt{fmul} & SDC in \emph{edges} & 672/2000 (33.6\%)\\
\hline
\texttt{cmp} & SDC in \emph{edges} & 275/2000 (13.8\%)\\
\hline
\end{tabular}
\normalsize
\vspace{4mm}
\end{table}

Tables \ref{failure_rate_hydro}, \ref{failure_rate_mhydro}, \ref{failure_rate_mish}, and \ref{failure_rate_umma} individually list detailed failure rates of the four mini-applications from FleCSALE and CODY in our fault injection experiments. Generally, injecting register-level faults in \texttt{fadd} and \texttt{fmul} instructions of all benchmarks leads to considerable amount of contaminated values of the calculated local/global parameters such as \emph{fluxes} and \emph{momentum}. Such failure rates go up to 77.2\% (\texttt{fmul} in \emph{maire\_hydro}), and on average 48.8\% for \texttt{fadd} and 56.5\% for \texttt{fmul}. Corrupting \texttt{cmp} instruction causes SDC as well (up to 55.5\% and 34.7\% on average) except for \emph{maire\_hydro} where crashes occurred instead. Moreover, corrupting \texttt{imul} instructions of \emph{maire\_hydro} causes abnormal \emph{energy} values which fails assertions in the program, while for \emph{mish} it leads to crashes (\emph{hydro} and \emph{umma} do not use \texttt{imul} instructions). Within the 2000 runs with fault injection into each instruction type, we observe significantly high failure rates, which indicate that the compute-intensive functions of all benchmarks we are interested in are prone to register-level soft errors, and eventually different types of failures were triggered.

\subsubsection{Visualized Fault Injection Results}

\begin{figure}[h]
\center
\begin{tabular}{cc}
\hspace{-0.6em}\includegraphics[width=1.6in]{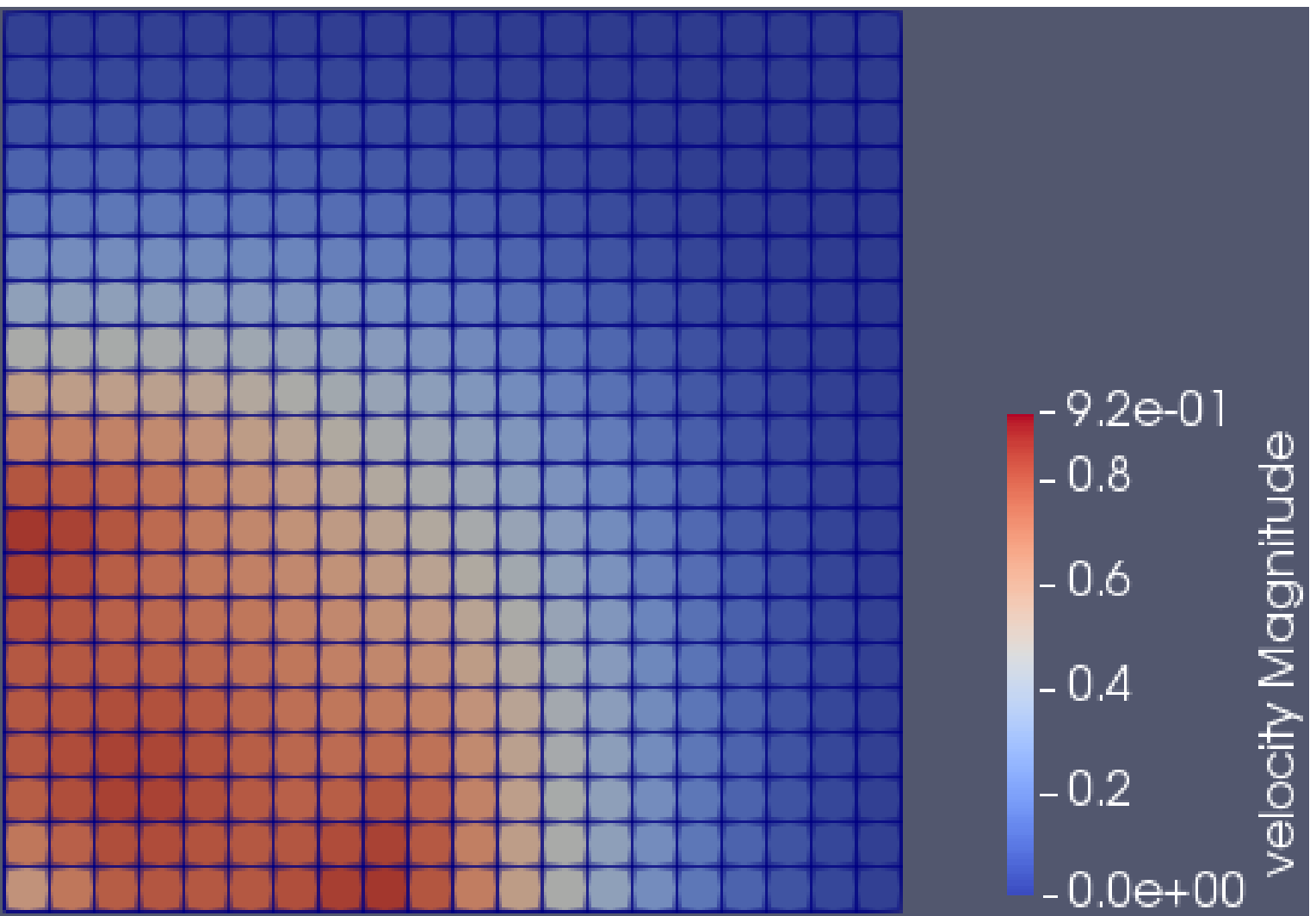} & \includegraphics[width=1.6in]{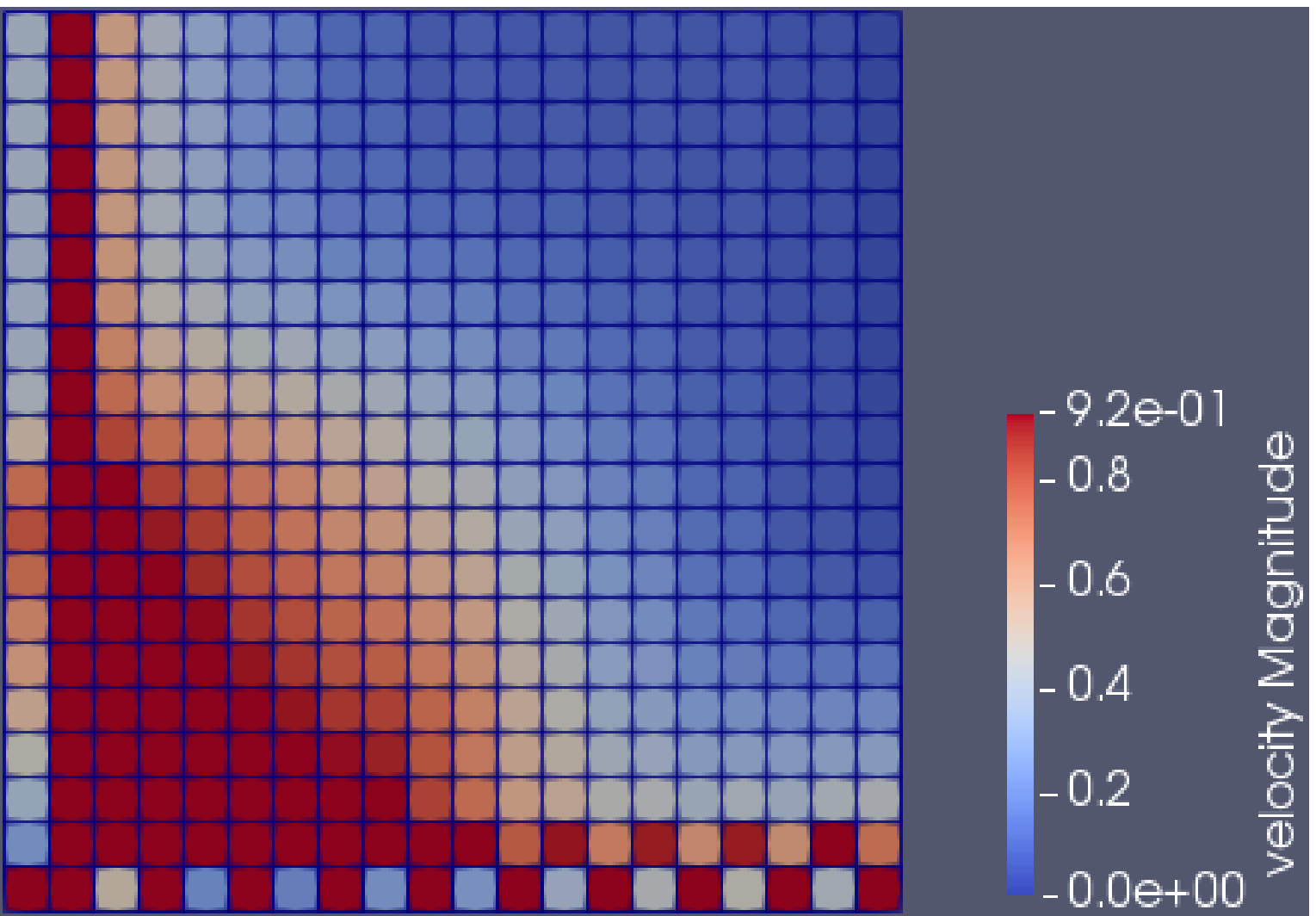}\\
(a) Original (velocity) & (b) Faulty (velocity)\\
\end{tabular}
\caption{Visualized Outputs of Correct/Faulty Runs of \emph{hydro}.}
\label{original_faulty_hydro3d}
\end{figure}

\begin{figure}[h]
\center
\begin{tabular}{cc}
\hspace{-0.6em}\includegraphics[width=1.6in]{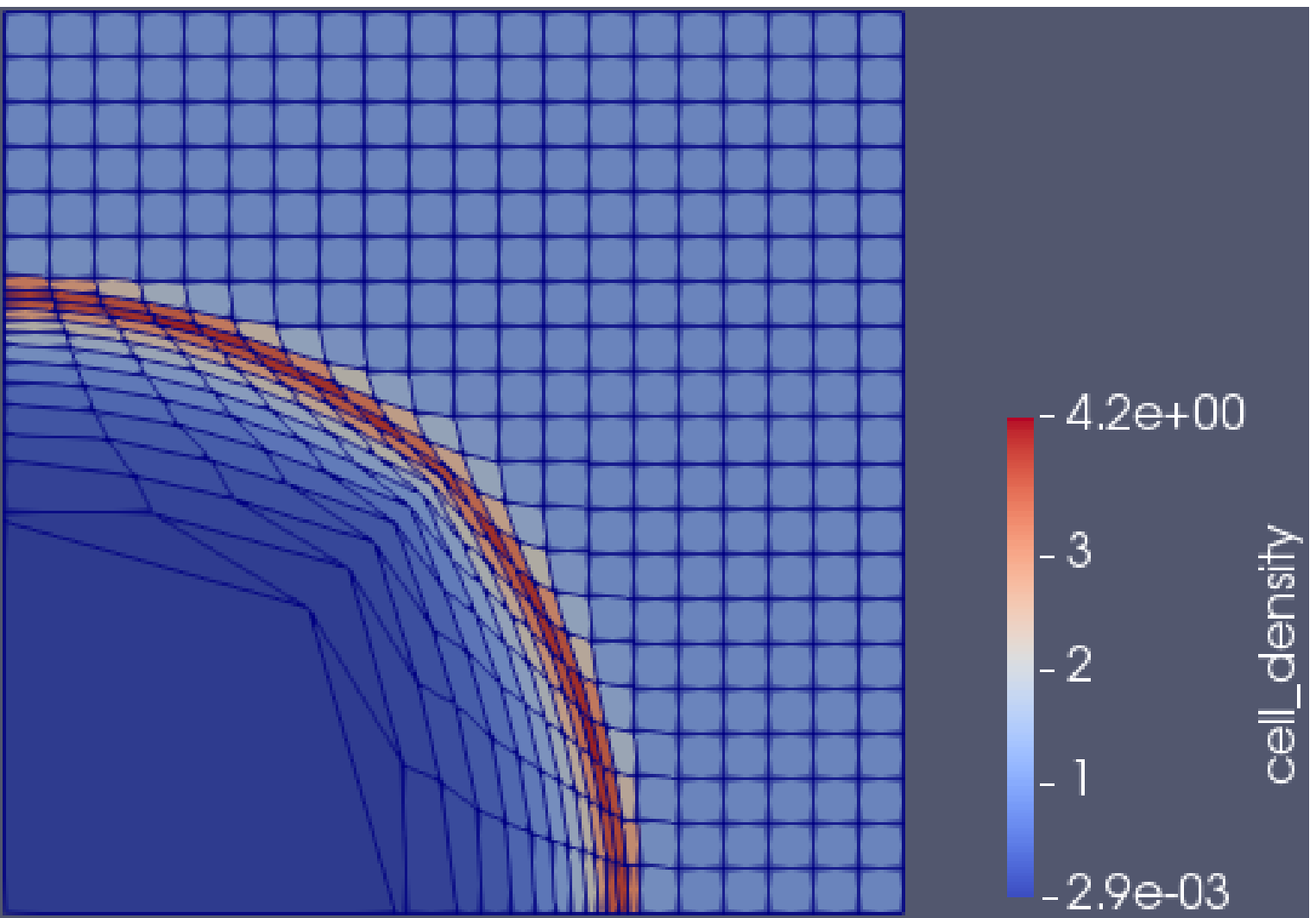} & \includegraphics[width=1.6in]{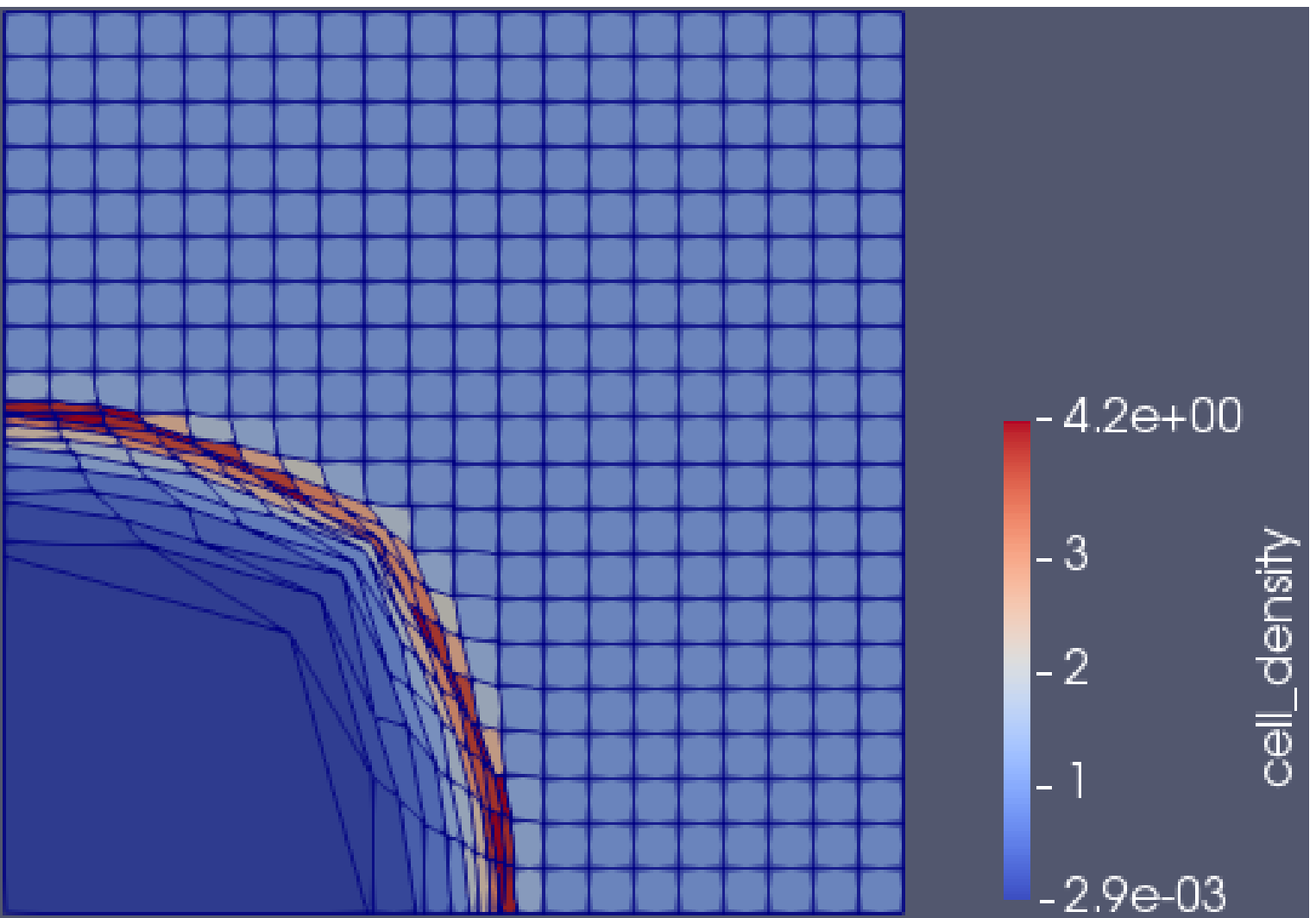}\\
(a) Original (cell density) & (b) Faulty (cell density)\\
\end{tabular}
\caption{Visualized Outputs of Correct/Faulty Runs of \emph{maire\_hydro}.}
\label{original_faulty_mhydro3d}
\end{figure}

\begin{figure}[h]
\center
\begin{tabular}{cc}
\hspace{-0.6em}\includegraphics[width=1.6in]{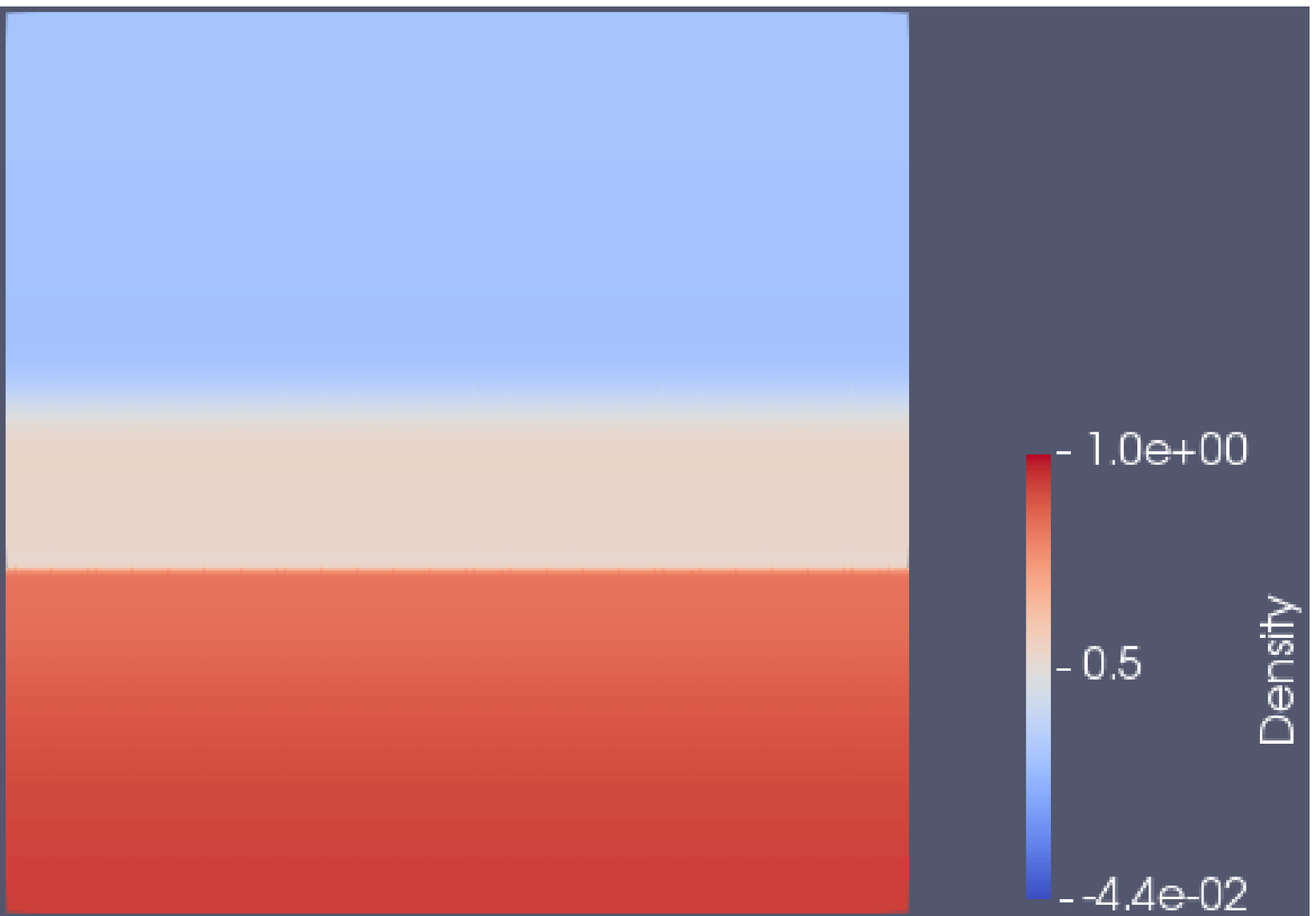} & \includegraphics[width=1.6in]{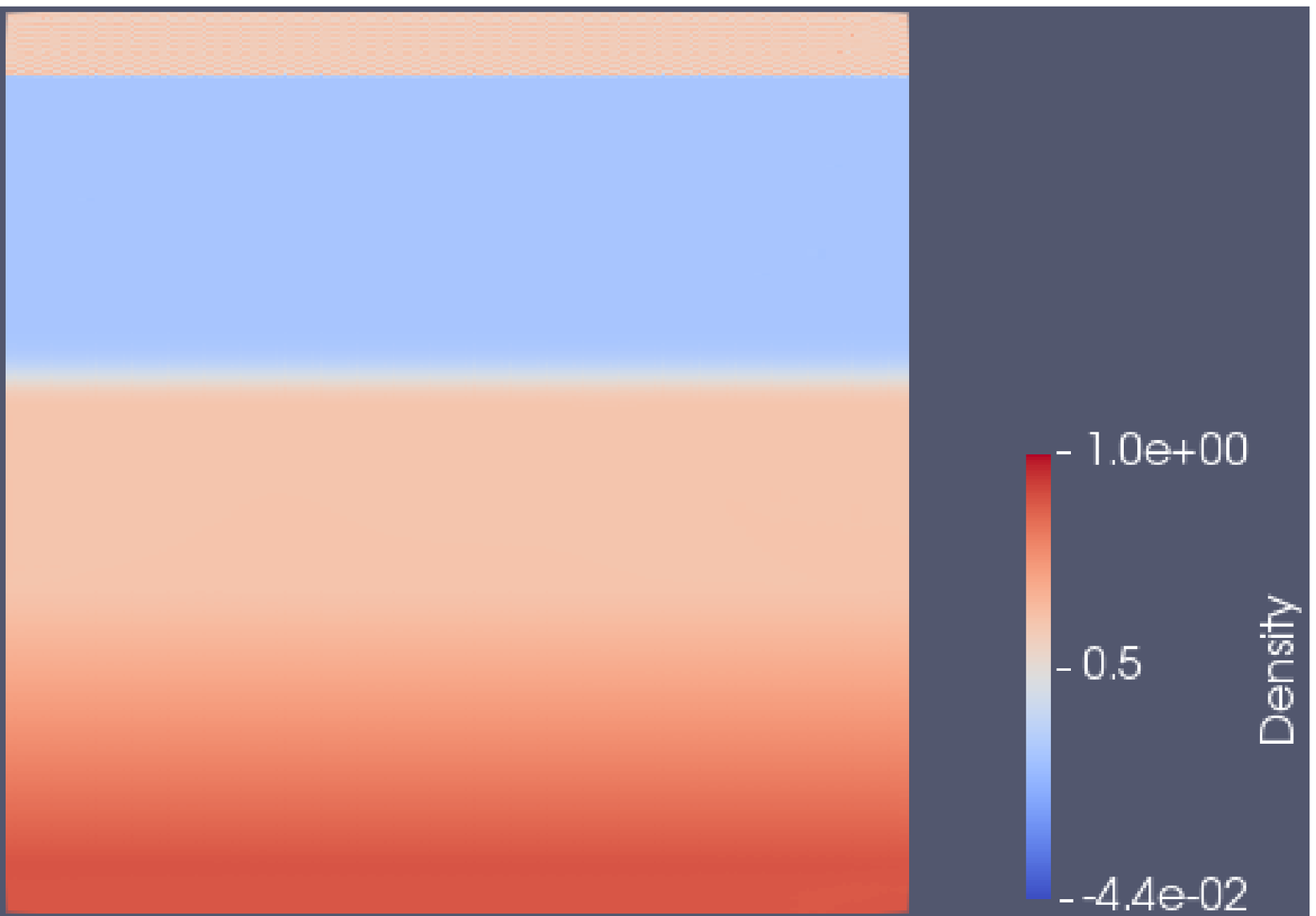}\\
(a) Original (density) & (b) Faulty (density)\\
\end{tabular}
\caption{Visualized Outputs of Correct/Faulty Runs of \emph{mish}.}
\label{original_faulty_mish}
\end{figure}

For better understanding of failure propagation in target application runs, we demonstrate the visualized program outputs with and without fault injection individually (two runs using the same initial parameters for input size) for \emph{hydro} (Figure \ref{original_faulty_hydro3d}), \emph{maire\_hydro} (Figure \ref{original_faulty_mhydro3d}), and \emph{mish} (Figure \ref{original_faulty_mish}). We did not present the results of \emph{umma} because this program does not produce outputs that are visualizable. Specifically, the snapshots shows final values (at a certain time step) on particular program variables across the mesh, e.g., velocity for \emph{hydro}, and cell density for \emph{maire\_hydro}. It is observable that significant differences exist from correct runs to faulty runs. For example, the snapshots in Figure \ref{original_faulty_hydro3d} indicate that the injected faulty values in one cell spread into adjacent cells. The snapshots in Figure \ref{original_faulty_mhydro3d} show the variation of cell density in \emph{maire\_hydro} after building the mesh. We can see the the major differences are: (a) the depth of growing (the original run spreads further than the faulty one), and (b) the cell density difference (the corrupted values in the faulty run are greater than the original).

Figure \ref{original_faulty_mish} compares the original runs of \emph{mish} without fault injection with the faulty runs. There exists significant difference in the gradient and the values of density as the generated mesh grows. Specifically, in the faulty runs, entities of the mesh have less density with slower varying gradient, since the fault injection corrupts key parameter values during the computation which affects the values of density and gradient from iteration to iteration. Note that these snapshots demonstrate the SDC scenarios only.

\subsection{Failure Recovery}

\subsubsection{Failure Recovery Coverage}

Experimental results show that our checksum-retry approach covers all SDC, i.e., all injected faults that lead to SDC were detected and recovered online (we only injected one fault in one run). However, the proposed checksum-retry cannot recover runs from crashes (e.g., segmentation fault in our experiments) online, since the nature of this type of failures is beyond the detection capability of our approach. For the assertion failures based on conservation laws, we re-executed the computation from the last correct iteration, similarly to the demonstrated checksum-retry approach.

\subsubsection{Performance Loss}

\begin{figure}[h]
\centering
\includegraphics[width=3.27in]{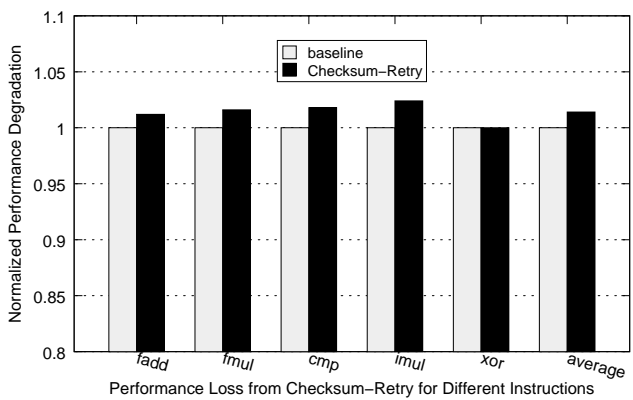}
\caption{Performance Loss from Checksum-Retry for Failure Detection and Recovery.}
\label{perf_loss}
\end{figure}

Figure \ref{perf_loss} compares the performance of original runs of all benchmarks, and runs with the checksum-retry approach for failure recovery from fault injected into different instruction types. All performance data shown is average across all runs and normalized to the original runs. We can see that our checksum-retry approach can successfully detect and recover from injected errors (SDC), while incurring negligible performance loss up to 2.4\% and on average 1.4\%. Compared to traditional resilience techniques C/R, it is lightweight in both storage usage and memory accesses. Only the last correct application state needs to be saved in memory for failure recovery if error occurs, and the same memory space keeps overwriting itself without new memory needed to allocate. Note that for \texttt{xor} instruction, the fault injection did not bring any observable errors, and thus experimentally the checksum-retry functionality was not actually employed in this case. Excluding this corner case leads to an average performance loss of 1.75\% from our checksum-retry approach when failures are triggered.

\section{Related Work}

Application resilience today is a key challenge and of great concern by the HPC community. Various resilience efforts have been conducted towards scientific applications of different domains. Gamell \textit{et al.} \cite{sc14} studied the efficiency of recovery from process/node failures for MPI applications in an online and transparent manner, with the help of application-driven, diskless, implicitly-coordinated checkpointing. The subsequent work \cite{sc15} explored the practice of enabling online and transparent local recovery for stencil-based parallel applications running at extremely large scale. They discussed how to mask multiple independent failures for overhead reduction and implemented a runtime FenixLR of local recovery algorithms. Experimental results on a Titan Cray XK7 supercomputer demonstrated the sustained performance and scalability with frequent real node failures. In this work, we also study failure detection and recovery for scientific applications of specific domains, but we focus more on utilizing the application characteristics for a better resilience solution. Ropars \textit{et al.} \cite{sc13} proposed a different approach from traditional checkpointing protocols by identifying a common property and a partial-order relation of MPI applications. Leveraging the two notations, they devised an approach combining a hierarchical coordinated checkpointing and message logging, without penalizing recovery performance. Unlike this generic approach using checkpointing, we develop a checksum-retry resilience technique based on application-specific knowledge at library level.

Kim \cite{ipdps14b} presented a quantitative study on the impact of transient errors in GPU on global properties of the N-body simulation, such as the total energy discussed in our work. They proposed an error detection technique for N-body programs by utilizing two types of properties that are known to hold in the simulated physical models, similarly to the heuristics of invariants in our approach. Their approach focused on GPU devices while our work was only evaluated on CPU. Tan \textit{et al.} \cite{radiance17} discussed the validity of failure rates due to register vulnerability on a wide scope of scientific applications, while our work does not need to consider since we only collect failure data based on program outputs -- the invalidated errors in fault injection are not included in our resilience analysis. Guan \textit{et al.} \cite{cluster15} investigated the resilience to soft errors on a hydrodynamics mini-application, and designed a fault detection method using checkpoint/restart. Their approach was evaluated to incur minimal overhead (less than 1\%) with a good error detection rate (88.3\%). Similarly, our approach has negligible performance overhead.

\section{Conclusions and Future Work}

With the motivation of improving application resilience at scale to realize the exascale vision, we investigate the susceptibility to soft errors online of several DOE-concerned continuum dynamics software packages, FleCSALE and MISH, using the open-source fault injector F-SEFI. Initial experimental results show that continuum dynamics applications are prone to register-level faults in different types of instructions of interested compute-intensive functions. Therefore, we propose to leverage the application nature of complying with domain conservation laws to detect errors, and devise an invariant-based checksum-retry approach to detect and recover from failures online. Experimental results with extensive fault injection using F-SEFI on a virtualized platform demonstrate the effectiveness and efficiency of our approach.

Regarding paths forward, we plan to fulfill the following tasks: (a) evaluate the resilience of our approach for more applications sharing similar program characteristics as FleCSALE and MISH (even applications from other domains), and (b) study the scalability of our approach by running large-scale experiments with parallel fault injection campaigns. Overall, we expect to extend the proposed approach to be applicable to more domain-specific applications with high scalability and efficiency.

\bibliography{elsarticle-template}

\end{document}